%% file: main.tex
\documentclass[conference, 9pt]{IEEEtran}
\usepackage{caption}
\usepackage{graphicx}
\usepackage{subcaption}
\usepackage{gensymb}
\usepackage{xcolor}
\usepackage[english]{babel}

\usepackage[binary-units=true]{siunitx}
\usepackage{glossaries}
\loadglsentries{abbreviations}

\usepackage[backend=biber,style=ieee,dashed=false]{biblatex }
\AtBeginBibliography{\footnotesize}

\usepackage{tabularx}
\newcolumntype{Y}{>{\centering\arraybackslash}X} 
\newcolumntype{L}{>{\raggedright\arraybackslash}X}
\newcolumntype{R}{>{\raggedleft\arraybackslash}X}
\usepackage{cellspace}
\usepackage{multicol}
\usepackage{multirow}
\usepackage{tablefootnote}

\usepackage{url}

\usepackage[super]{nth}

\usepackage{tikz}
\usetikzlibrary{patterns}
\usetikzlibrary{shapes,arrows}
\usetikzlibrary{shapes.multipart}
\usetikzlibrary{arrows.meta}
\usetikzlibrary{circuits.ee.IEC}
\usepackage[americaninductors]{circuitikz}

\usepackage{pgfplots}
\usepgfplotslibrary{groupplots, dateplot}
\pgfplotsset{compat=newest}

\usepackage{siunitx}

\begin{document}
\title{Techtile: a Flexible Testbed for Distributed Acoustic Indoor Positioning and Sensing\vspace{-1cm}}


\author{\IEEEauthorblockN{Daan Delabie, Bert Cox, Lieven De Strycker, Liesbet Van der Perre}
\IEEEauthorblockA{\IEEEauthorblockA{
        KU Leuven, WaveCore, Department of Electrical Engineering (ESAT), Ghent Technology Campus\\
        B-9000 Ghent, Belgium\\
        daan.delabie@kuleuven.be
    }}}

\maketitle

\begin{abstract}
The proposed infrastructure, named Techtile, provides a unique R\&D facility as features dispersed electronics enables transmission and capturing of a multitude of signals in 3D. Specific available equipment that enhances the design process from smooth prototyping to a commercial product is discussed. The acoustic parameters of the room, particularly the reverberation and ambient noise, are measured to take these into account for future innovative acoustic indoor positioning and sensing systems. This can have a positive influence on the accuracy and precision. The wooden construction represents an acoustically challenging room for audible sound with a maximum measured RT60 value of 1.17s at 5kHz, which is comparable to the reverberation properties of a lecture or small concert hall. For ultrasound it is rather challenging due to the present ambient noise sources. In general, the Techtile room can be compared with a home or quiet office environment, in terms of \glspl{spl}. In addition to the acoustic properties, possible research and development options are discussed in combination with the associated challenges. Many of the designs described are available through open source.
\end{abstract}


\glsresetall


\section{Introduction}
A great demand for indoor wireless localization and sensing applications is experienced in healthcare, assisted living, home automation, industry 4.0 and surveillance and security systems. This requires a flexible testbed that enables fast development and comparability of different technologies \cite{survey_paper}. For indoor positioning, acoustic signals are interesting candidates given the relatively slow propagation speed of sound, the low frequency components that result in lower required sample rates, and the cheaper hardware combined with recent component developments such as \gls{mems}. Furthermore, the synchronization accuracy is less strict compared to \gls{rf}-based systems. Nevertheless, acoustic signals are subject to room characteristics and environmental factors such as reverberation, relative air velocity, ambient noise and temperature. 

\begin{figure}[!htb]
\centering
    \centering
    \includegraphics[width=\linewidth]{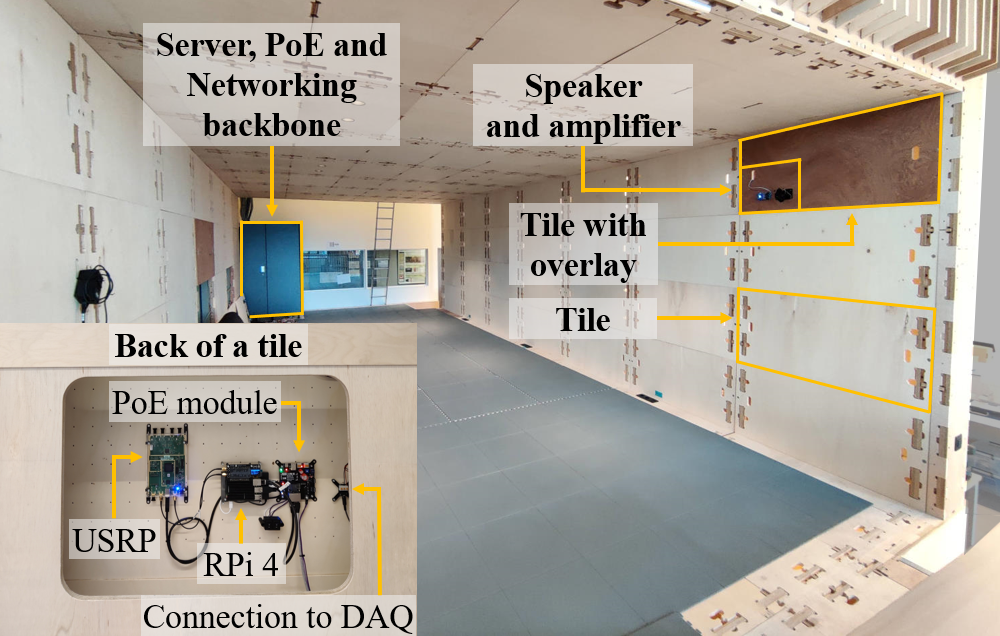}%
    \caption{\small The Techtile infrastructure with 140 equipped tiles. Each tile has a \gls{usrp} NI B210, RPi 4, \gls{poe} module and connection to the \gls{daq} which is standing next to Techile and therefore not in the picture.}%
    \label{fig:techtile}
\end{figure}

Only few  testbeds for acoustic-based indoor positioning and acoustic sensing have been reported so far, in comparison to testbeds for \gls{rf} research. An anechoic chamber is often used for sound measurements, in which ambient noise and reflections are reduced to a minimum. However, this is not suitable for testing real-life situations, which is our aim with the proposed Techtile testbed, depicted in Fig.~\ref{fig:techtile}, which integrates distributed electronics in 3D offering an unprecedented spatial diversity. Most real-life acoustic testbeds consist of separate nodes that are placed in a room~\cite{arabis, implement, doppler, practical_impl}. Some well-known examples are Cricket, BUZZ and Dolphin \cite{review}, specifically designed for indoor positioning. A dedicated acoustic testbed for small-scale areas (\SI{50}{c\meter} x \SI{50}{c\meter}) is also proposed and consists of a wooden base with 4 microphone modules in the corners~\cite{small}. Besides indoor situations, underwater acoustic localization is often the main focus of the acoustic testbed and can be found both as a simulation model~\cite{underwatersim} and as an implemented setup in the sea~\cite{lobster}. None of the existing real-life testbeds mention acoustic properties of the environment or offer a flexible room-size test infrastructure, which can simplify the repeatability of measurements. Furthermore, the focus is almost exclusively on indoor positioning, while sensing is only applied in already existing environments, e.g. a house, in which little or no attention is paid to the properties of the room. In addition, a limited number of microphones or speakers are usually used, which does not benefit the spatial diversity.

In contrast to other testbeds, Techtile is a dedicated measurement infrastructure that offers the possibility to implement a lot of microphones, speakers and other equipment in a flexible way. The option is provided to integrate these devices into the walls, ceiling or floor. Furthermore, furniture, obstacles, absorbers or additional noise sources can be added into the environment to test a wide set of situations and \gls{nlos} conditions. The acoustic parameters of this real-life, acoustically challenging chamber are measured, which is not the case for other testbeds. Provided that these parameters are known, a simulation model can be drawn up more correctly, after which practical tests are possible for validation. 
This testbed also offers other R\&D opportunities for communications, \gls{wpt}, general indoor positioning via \gls{rf}, light, IR, etc. and combinations or comparisons with, among others, acoustic setups \cite{radioweavesPaper, olderTechtilePaper, newTechtilePaper}. However, the focus of this paper is on the acoustic properties of this testbed and the available equipment needed for future research on acoustic positioning and sensing.

The paper is structured as follows. First, the Techtile infrastructure with the available equipment that can be used for acoustic positioning and sensing is discussed in Section~\ref{sec:technical}. These technical characteristics are important to frame future possibilities. Subsequently, the acoustic characteristics of the room are explained in Section~\ref{sec:char}. With this knowledge, acoustic research and development, discussed in Section~\ref{sec:development}, can be elaborated. Lastly, we present potential challenges in Section~\ref{sec:challenges} and a conclusion in Section~\ref{sec:concl}.

\section{Techtile Infrastructure and Technological Features}
\label{sec:technical}
Techtile, the WikiHouse inspired room inside a room, is a \SI{8}{\meter} x \SI{4}{\meter} x \SI{2.4}{\meter} multi-functional measurement infrastructure, built for easy deployment of distributed sensing and communication technologies. The composition of 140 equal sized tiles covering the walls, ceiling and floor creates a versatile and adaptive 3D environment, depicted in Fig.~\ref{fig:techtile}. Each tile is equipped with a \gls{poe} module, a Raspberry Pi 4 edge computing device, and a \gls{sdr}. These are mounted on the back of the tiles, facilitating hidden electronics from inside the room's viewpoint. Connection to the processing backbone is provided through high throughput Ethernet \cite{olderTechtilePaper, newTechtilePaper}. For acoustic sensing and positioning, hiding the acoustic hardware behind the tiles is unfavorable, as most of the signals get reflected and do not enter the infrastructure. The tiles are for this reason extended on the front side with an overlay panel, enabling easy mounting of microphones or speakers. An inherently synchronized connection of the \gls{daq} to each sensor (e.g. microphone) or actuator (e.g. speaker) enables fast prototyping.

\begin{figure}[!htb]
\centering
    \centering
    \includegraphics[width=\linewidth]{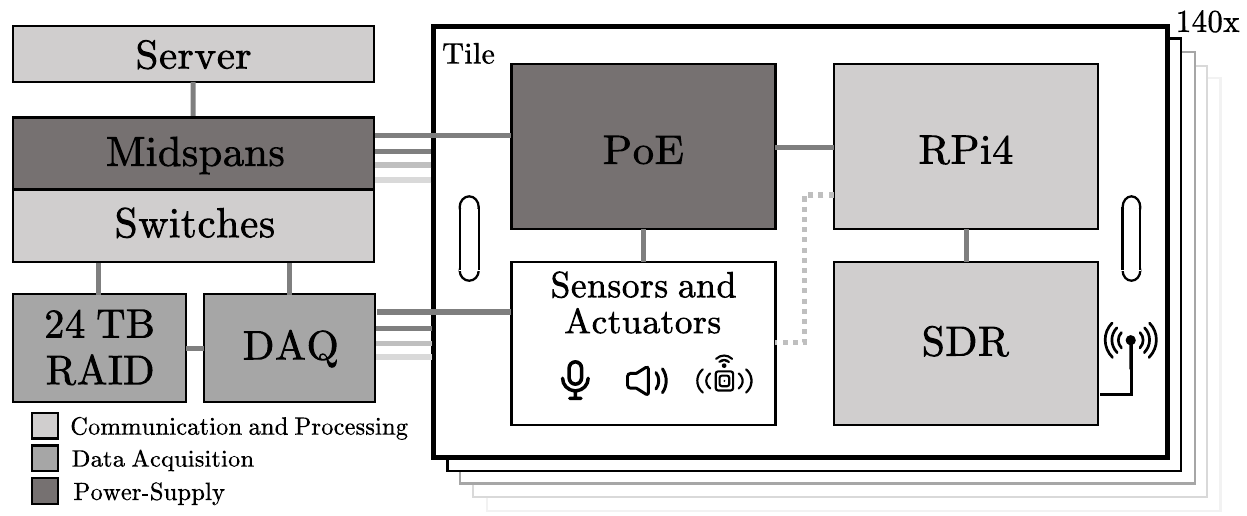}%
    \caption{\small Overall system diagram of the Techtile infrastructure, related to communication and processing, power-supply, \gls{daq}, and acoustic sensors and actuators.}%
    \label{fig:archtiectureOverview}
\end{figure}

The following subsections focus on the Techtile building blocks, depicted in Fig.~\ref{fig:archtiectureOverview}, that are important for acoustic positioning and sensing. For other applications, the correlated technical aspects and potential research contribution of this infrastructure, we kindly refer to the following research \cite{olderTechtilePaper, newTechtilePaper}.

\subsection{Acoustic sensors and actuators}

\subsubsection{MEMS Microphone}
\begin{figure}[!htb]
\centering
\begin{subfigure}[b]{.18\textwidth}
  \centering
  \includegraphics[width=0.85\linewidth]{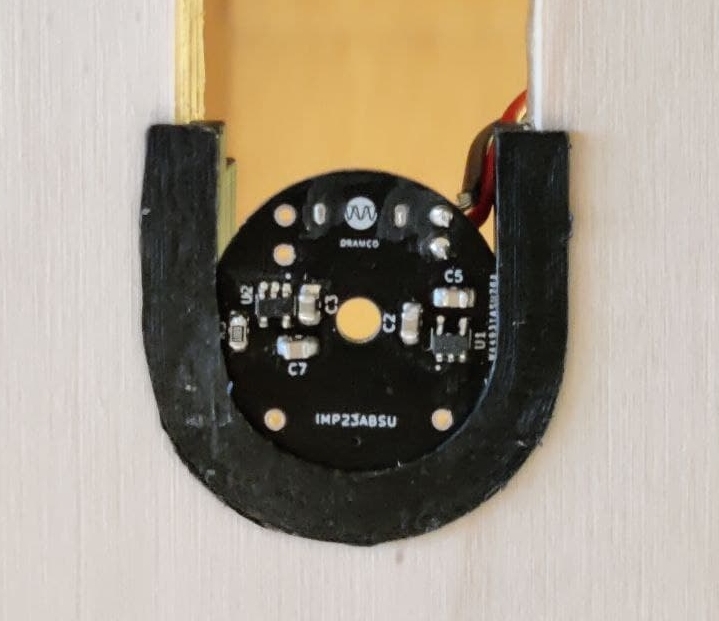}
  \caption{}
  \label{fig:sub1}
\end{subfigure}%
\begin{subfigure}[b]{.32\textwidth}
  \centering
  \includegraphics[width=0.85\linewidth]{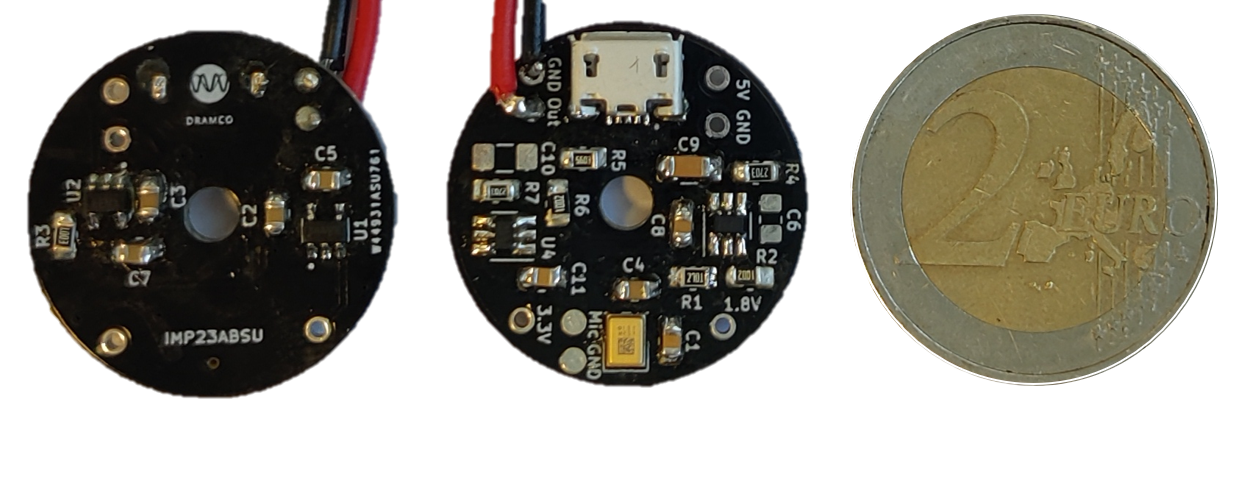}
  \caption{}
  \label{fig:sub2}
\end{subfigure}
\caption{\small \gls{mems} microphone module integrated in a tile (a) and close-up of the front and back of the hardware (b).}
\label{fig:MEMSmic_integration}
\end{figure}

The distribution of in-house developed microphone modules over the entire room contributes to the desired spatial diversity of the measurement infrastructure. Today we deployed 11 modules, yet this can easily be extended. The microphone modules are mounted in the handles of the tiles via a 3D printed element, as depicted in Fig.~\ref{fig:sub1}. This makes this setup flexible whilst maintaining the integrated electronics principle of this infrastructure.


The modules are designed with a \gls{mems} microphone (IMP23ABSU \cite{IMP23ABSU_datasheet}) since this offers a lower cost, more energy efficient and smaller option than a conventional microphone but still has good or even better properties such as bandwidth and sensitivity. This particular microphone has a bandwidth of \SI{80}{k\hertz} and a sensitivity of \SI{-38}{\decibel V} $\pm$ \SI{1}{\decibel} at \SI{94}{\decibel SPL}, \SI{1}{k\hertz}. The received acoustic signal is amplified in two stages with low-power, rail-to-rail TLV341 OPAMPs. The typical gain bandwidth product of this OPAMP is specified as \SI{2.3}{M\hertz}. If a bandwidth of, for example, \SI{80}{k\hertz} is desired, a maximum gain of \SI{57.9}{\decibel} can be set via the 2 amplification stages. However, the first order filters are easy to adjust according to the desired audible or ultrasonic application.

\subsubsection{Speakers}
Similar to the microphones, speaker selection is application dependent. In this research we define two types: speakers for applications in the audible frequency domain, and speakers for ultrasonic purposes. The first category is sufficiently supported by commercially available speakers (e.g. HiFi, computer or smartphone speakers) as long as the bandwidth of these integrated speakers lies between \SI{20}{\hertz} - \SI{20}{k\hertz}. Typical applications based on audible sound, are presence detection, acoustic sensing, mapping, noise pollution control, sound source localization and sound recognition for e.g. assisted living.
Unlike audible sounds, ultrasonic signals are not intrusive and for this reason often used for indoor positioning. Dedicated, typically piezo based, speakers are needed. Within the Techtile environment, a number of Kemo L010 ultrasonic speakers with a frequency range of \SI{2}{k\hertz} - \SI{60}{k\hertz} combined with off-the-shelf amplifiers with a frequency range over 45 kHz are mounted on the tile's overlays. These speakers are currently used for \gls{rf}-acoustic indoor positioning of energy neutral devices~\cite{IPIN2021, asilomarBert}. Vice versa, if an ultrasonic speaker based mobile node needs to be localized, a more compact, energy-efficient and omnidirectional speaker is desired, which is not the case with the Kemo L010. Again \gls{mems} technology offers the solution with the added advantage that these speakers are lower cost. In Techtile, the ADAP UT-P2019 \gls{mems} speaker is integrated in a mobile node, accompanied with the associated LM48580 amplifier. Typically, the output \gls{spl} level of \gls{mems} speakers is lower than traditional speakers, forming a vast challenge for positioning purposes.

\subsection{Data Acquisition}
The \gls{daq} provides simultaneous sampling of input and output signals through several parallel channels and is the main backbone element to validate theoretical results on acoustic indoor positioning and sensing with practical experiments. Favorable ideal circumstances can be extended to more realistic scenarios with loosely coupled configurations using lower cost edge devices and in-house designed hardware. The \gls{daq} consists of an 18-Slot PXI Express chassis (NI PXIe-1095), containing an NI PXIe-8080 controller, NI PXIe-8394 bus extension module used as connection to a 24TB external RAID (RMX-8268), NI PXIe-6672 timing and synchronization module and 12 NI PXIe-6358 multi-function I/O modules. These I/O modules are in turn connected to 24 SCB-68A noise rejecting, shielded I/O connector blocks. Through these blocks, user-friendly connections via screw terminals to a total of 384 single-ended or 192 differential analog input channels, 48 analog output channels, and 576 digital I/O channels are provided. The large amount of available synchronized channels offers the possibility to realize large distributed sensor array setups evolving into the presence of a sensor such as a microphone on each tile. 

The massive \gls{daq} provides a 16-bit \gls{adc} and \gls{dac} resolution with a maximum sample rate of \SI{1.25}{M\hertz} and \SI{3.3}{M\hertz} for respectively the analog input and output channels. For sound and even ultrasonic systems (in our research currently up to \SI{45}{k\hertz}) this sample rate is more than sufficient and certainly satisfies the Nyquist theorem. Assuming that the speed of sound ($v_{sound}$) is equal to \SI{343}{\meter/\second} and sampling ($f_s$) is done at the highest possible input sample rate, a theoretical minimum ranging resolution per sample ($e_{min}$) of \SI{0.27}{m\meter} is found for indoor time based positioning. As a result, quasi-optimal conditions are created to minimize errors.

\begin{equation}
    e_{min} = \frac{v_{sound}}{f_s} = \frac{343 \, \textrm{m/s}}{1.25 \, \textrm{MHz}} = 0.27 \, \textrm{mm}
    \label{eq:distanceErrorTh}
\end{equation}

For the ordinate, the analog output voltage range can be set to $\pm 10 \rm\,V$ and $\pm 5\rm\,V$ while for the analog input the range is extended to $\pm 1\rm\,V$ and $\pm 2 \rm \,V$. Given the constant 16-bit sample resolution, different \gls{fs} values result in different absolute accuracy, as presented in Table~\ref{tab:resolutionDaQ}.

\input{tables/accuracy_daq}

The timing and synchronization controller synchronizes the different modules within the same chassis to support high-channel-count applications, which can even be expanded for multi-chassis purposes. A high-stability \SI{10}{M\hertz} timebase with \SI{1}{ppm} accuracy and a \gls{dds} clock from DC up to \SI{105}{M\hertz} with a resolution less than \SI{0.075}{\hertz} is provided. 

\subsection{Power-Supply}
In Techtile, a number of options for power supply are provided. In addition to the traditional mains power supply that is present in a number of fixed places, a \gls{poe} connection is available on every tile, delivering the desired DC-voltages. The \gls{poe} technology presents the most interesting power source since ideally all tiles only need a single Ethernet connection for power, communication and synchronization. The in-house designed \gls{poe} modules can deliver a maximum power of \SI{90}{\watt} each and are supported by several \gls{poe} midspans delivering a maximum total power of \SI{9}{k\watt}. This is typically sufficient to feed acoustic test infrastructures. In addition, the Techtile environment is used for research related to wireless power transfer for both far-field and near-field applications.

\subsection{Communication and Processing}
Each tile is connected via Ethernet to a central network cabinet consisting of several \SI{10}{Gbit} Dell S4148T-ON switches. In addition, a connected central server (Dell PowerEdge R7525) can handle simulations and calculations with the data in order to process it. On each tile, a RPi 4 is available that can serve as a local processing unit to enable edge computing. The RPi is equipped with 8GB SDRAM and a Quad core Cortex-A72 64-bit \gls{soc}, and can in turn control the \glspl{sdr} (NI B210) that are also present on each tile. Via these \glspl{sdr} and associated antennas, wireless communication, synchronization and charging can be provided for mobile nodes within Techtile. 
As a result, hybrid \gls{rf}-acoustic indoor positioning within the measurement infrastructure is made possible. 

\section{Acoustic Characteristics}
\label{sec:char}
In order to perform reliable measurements and to compare the results with other set-ups, we conducted several measurements to characterize the acoustic properties of the infrastructure. This paper focuses on \gls{rt} and background noise since these parameters are widely used to define and evaluate the acoustic performance of a room and these have a significant impact on indoor positioning and sensing systems.

\subsection{Reverberation Time}
Typically, the reverberation of a chamber is described using the RT60 value. According to the ISO 3382-3:2022 \cite{ISOISO3367}, specifying a measuring method for acoustic space parameters in unoccupied offices, the RT60 in an enclosure is defined as the required duration for the space-averaged sound energy density to decrease by \SI{60}{\decibel} after interrupting the sound source. The measurements are performed using the NTi XL2~\cite{datasheetNTi} with an M2211 microphone, performing a Schroeder-method~\cite{schroeder} RT60 measurement based on an extrapolated RT20 value. The commonly chosen method using a popping balloon as an impulse for the measurements~\cite{balloon_popping} was applied in a mostly empty Techtile environment. The Techtile room contained a fixed cloth seat. The results can be found in Table~\ref{tab:rt60}. 
Due to the limitations of the measuring device, no specific ultrasonic measurements could be performed. RT60 values in the ultrasonic domain are expected to be lower as the reverberation time is affected by the frequency-dependent absorption of sound through air which increases at higher frequencies~\cite{atmo_abs}.

\input{tables/rt60_table.tex}

Based on these RT60 measurement values it could be concluded that the Techtile room behaves as a real-life, acoustically acceptable environment which can be compared to a classroom, office or lecture hall\footnote{\url{https://www.nti-audio.com/en/applications/room-building-acoustics/reverberation-time}}. Nevertheless the critical distance $d_c$, being the distance at which the direct sound and reverberated sound can exhibit equal \gls{spl} values, at \SI{500}{\hertz} is equal to \SI{0.46}{\meter}. This is rather a small value since the Techtile space is typically smaller (\SI{77}{\meter^3}) than the listed examples.
The formula for determining this distance (\ref{eq:critdist}) \cite{2005Tad} is based on the Sabine's reverberation formula given in (\ref{eq:sabines})~\cite{sabine}, with $\gamma$ the degree of directivity of the source, $V$ the room volume in $\rm m^3$, $C_{20}$ the speed of sound at \SI{20}{\degree C}, and $S_a$ the equivalent absorption surface in $\rm m^2$.

\begin{equation}
    d_c = 0.057 \sqrt{\frac{\gamma V}{RT_{60}}}
    \label{eq:critdist}
\end{equation}

\begin{equation}
    RT_{60} = \frac{24 \, \textrm{ln}(10) V}{C_{20} S_a}
    \label{eq:sabines}
\end{equation}

The critical distance value is rather small, which means that large reverberated \gls{spl} levels are already perceptible from a particular small distance relative to the speaker. It can be concluded that, given the limited space, the Techtile room is an acoustically harsh environment.

\subsection{Ambient Noise}
\begin{figure}[h!]
\centering
    \centering
    \includegraphics[width=0.75\linewidth]{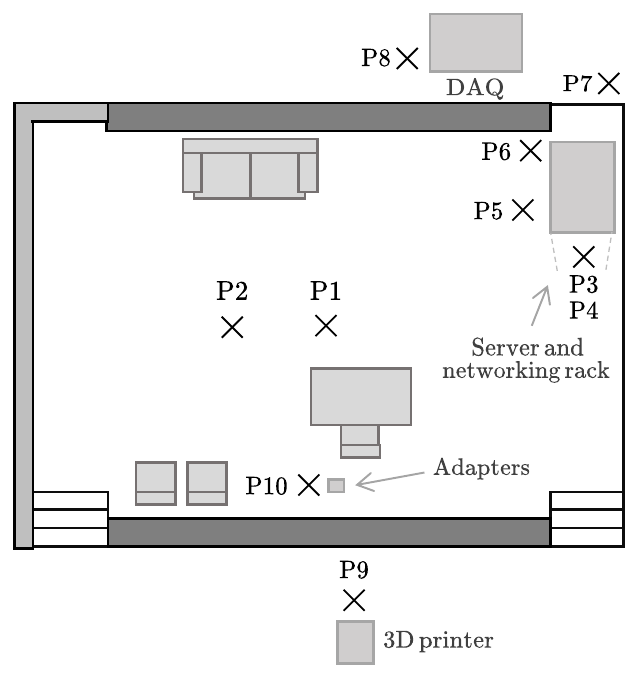}%
    \caption{\small Top view of Techtile showing the different \gls{rta} measure positions $\rm PX$.}%
    \label{fig:meas_locations}
\end{figure}

The Techtile infrastructure consists of embedded electronics, power, sampling and processing units, all contributing to the (audible) ambient noise. For future acoustic measurements, it is important to characterize their contribution to the noise-induced distortion as this could greatly impact the accuracy and reliability of these measurements.
Based on a \gls{rta}, measured via the REW V5.19 Room Acoustic Software, a Behringer U-PHORIA UMC202HD audio-interface and the previously discussed microphone module based on the IMP23ABSU, the frequency spectrum with the corresponding maximum \gls{spl} values are measured within a 10 second window. To reveal the different noise sources, measurements were taken at 10 different positions (P) as sketched in Fig.~\ref{fig:meas_locations}. The microphone was calibrated at \SI{1}{k\hertz} via the NTi XL2 and a sample rate of \SI{192}{k\hertz} was used.

\begin{figure}[h!]
\centering
    \centering
    \includegraphics[width=\linewidth]{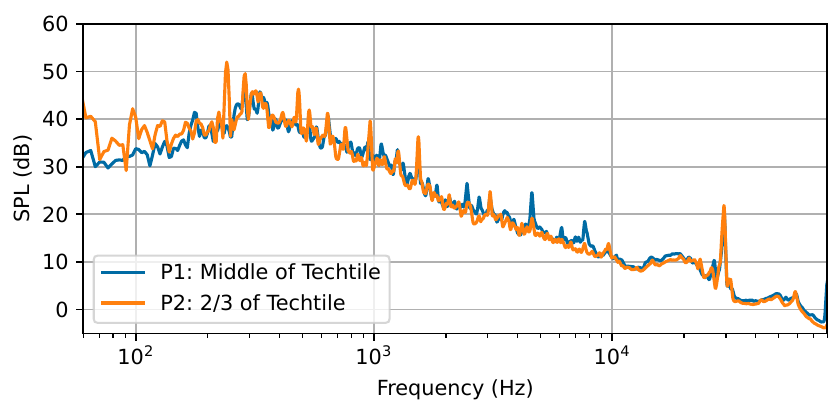}%
    \caption{\small Measured \gls{rta} in the Techtile environment.}%
    \label{fig:rta_meas_techtile}
\end{figure}

\begin{figure}[h!]
\centering
    \centering
    \includegraphics[width=\linewidth]{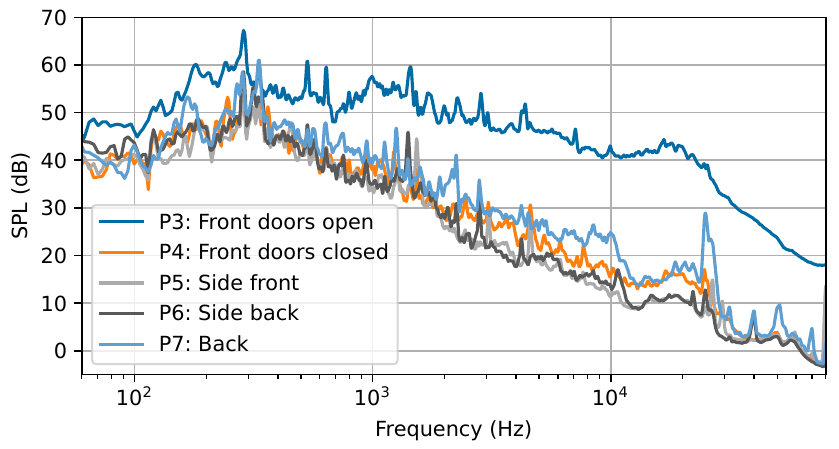}%
    \caption{\small Measured \gls{rta} near the server and networking rack.}%
    \label{fig:rta_meas_server}
\end{figure}

\begin{figure}[h!]
\centering
    \centering
    \includegraphics[width=\linewidth]{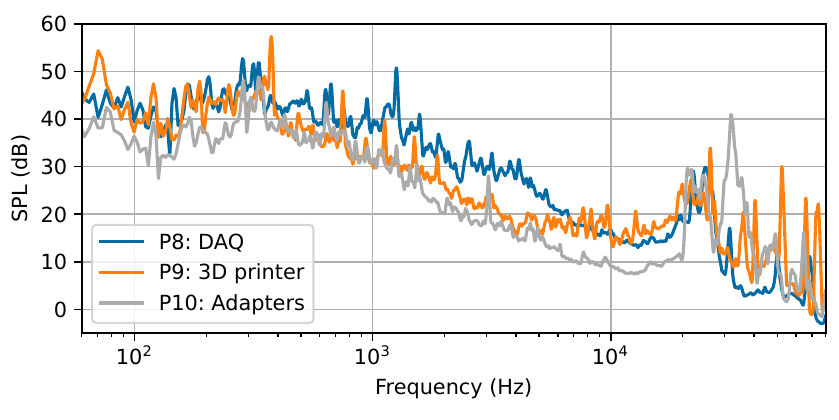}%
    \caption{\small Measured \gls{rta} coming from external devices near and in Techtile.}%
    \label{fig:rta_meas_noise}
\end{figure}

 The graphs in Fig.~\ref{fig:rta_meas_techtile}, Fig.~\ref{fig:rta_meas_server} and Fig.~\ref{fig:rta_meas_noise} show the recorded results respectively in the Techtile environment itself, around the server and networking rack and near other known sound-noise sources. Since the used microphone does not have an ideal flat response, and no compensation has been provided, these graphs are only valid for this specific microphone, which will be used during future research. Nevertheless, the results do give a good indication of where and at what frequency sources emit sound.

As shown in Fig.~\ref{fig:rta_meas_techtile}, the \gls{rta} results for P1 and P2 show mainly recorded sound in the audible range, with a peak value in the ultrasound range at \SI{29.5}{k\hertz}. When compared with measurements at the other positions (e.g. P4, P5, P8 and P10) it is clear that this is the result of a combination of the noise of the adapters located in the middle of Techtile, the noise of the server and networking backbone, and the \gls{daq}. In the middle of Techtile (P1) it is rather quiet as a maximum value of \SI{46.2}{\decibel} is measured. Even if flat response compensation of the microphone is applied according to the sensitivity given in the datasheet~\cite{IMP23ABSU_datasheet}, which is approximately the case up to \SI{10}{k\hertz}, this value corresponds with the noise in an average home or quiet office, which is an ideal situation for simulating realistic scenarios.

The noise reduction of the APC NetShelter CX rack, containing the server and networking infrastructure, is clearly visible when comparing the SPL values given in Fig~\ref{fig:rta_meas_server} at positions P3 until P7. High frequencies are more attenuated compared to low frequencies, with an average measured and certainly necessary noise reduction of \SI{22.9}{\decibel}. The noise reduction at the front of the rack is \SI{22.7}{\decibel} and is smaller than the reduction on the side (\SI{24.2}{\decibel}), yet significantly larger than at the back (\SI{20.4}{\decibel}). According to the datasheet~\cite{serverkast}, the noise reduction is \SI{18.5}{\decibel}, however no information is given about the used measuring methodology. Furthermore, these \gls{rta} measurements are snapshots in time, with the consequence that the data could not be gathered simultaneously with the current setup. Instead, one measurement after another was taken, which thus shows slightly different situations. In addition, it is stated in the datasheet that the cooling of the cabinet itself generates \SI{48.5}{\decibel} of noise during normal operation, which corresponds to the measured values in the audible range when the cabinet is closed.

Besides the server and networking infrastructure, the \gls{daq}, adapters and 3D printers nearby also provide clear noise components, as shown in Fig~\ref{fig:rta_meas_noise}. This noise is mainly located in the ultrasonic range. For example, a maximum peak of \SI{40.9}{\decibel}, coming from the adapters, can be observed at \SI{32}{k\hertz} while the 3D printer generates a maximum peak of \SI{33.9}{\decibel} at \SI{26.2}{k\hertz}. The speaker output values of the aforementioned ultrasonic localization systems should exceed these noise values, which is certainly possible since these measurements were performed right next to the device, while in the middle of Techtile (P1) this noise is rather limited. It is important to mention that the 3D printer only sporadically generates noise during printing as a result of which this is not visible in the measurements of Fig.~\ref{fig:rta_meas_techtile}.

\section{Acoustic Research and Development}
\label{sec:development}
Future measurements can take into account the knowledge of noise and reverberation, which can positively affect accuracy and precision. Some possible projects, which this infrastructure makes possible, are briefly discussed.

\subsection{Distributed Sensing}
A lot of distributed microphones can contribute to a more accurate localization and sensing system since there is more redundancy and reverberation effects can therefore be tackled. Using this setup, the influence of the number, selection, and position of the microphones on, for example, the accuracy of the position determination can be investigated. Two different setups with another nature of the mobile node can be distinguished: a low-power mobile speaker based mobile node and an ambient noise based mobile node. The latter option enables sound detection and classification for, e.g. assisted living. A performance assessment of an indoor positioning system requires a large number of spatial 3D samples. Since this involves a very intensive and time consuming measurement process, prone to human error, an automated rover~\cite{newTechtilePaper} was developed.

\subsubsection{Low-Power Mobile Speaker Based capturing}
This setup allows to determine the position of a mobile node consisting of a low-power (e.g. MEMS based) speaker module emitting an acoustic chirp signal in the ultrasonic region. Pulse compression allows \gls{tdoa} values to be obtained, which in turn can be used to extract 3D-positions. Since the microphone modules are all connected to the same \gls{daq}, synchronization is inherently present. Furthermore, this setup allows to experiment with distributed scheduling and different localization algorithms, such as physically inspired deep learning models.

\subsubsection{Ambient Noise Based capturing}
This setup is similar to the previous setup except that the speaker is replaced by sounds that occur naturally, such as people speaking, falling objects, footsteps, etc. This is better for the user's ease of use, but complicates the algorithms. The connection with acoustic sensing can lead to additional information related to the sound source.

\subsection{Energy-neutral Hybrid \gls{rf}-acoustic Localization}
Combining \gls{rf} and acoustics allows achieving the best of both worlds: the slower propagating acoustic signals for positioning purposes and the \gls{rf} signals for synchronization and communication. Mobile nodes in acoustic positioning often consist of microphone modules instead of speakers, as receiving requires less energy than transmitting acoustic signals. Energy-wise, adding \gls{rf} to the system is a dreadful decision but can be countered by adding \gls{rf}-energy harvesting hardware on the mobile node side~\cite{asilomarBert}. Another solution can be found by (almost) passive \gls{rf}-backscattering, where communication is acquired by either reflecting or absorbing the incoming \gls{rf} waves. This eliminates a power hungry \gls{rf} source at the node and enables, in combination with the energy harvesting, energy neutral positioning~\cite{IPIN2021}.
The Techtile infrastructure is the ideal environment for testing this hardware, as it houses \gls{usrp}'s for sending and receiving the \gls{rf} signals, raspberry Pi's for both processing and acoustic signal generation, and a modular tile composition for speaker positioning optimization.


\subsection{Acoustic SLAM}
Indoor positioning algorithms can benefit from environmental knowledge to improve the positioning performance. Installation and maintenance efforts are usually the bummer for not doing this. Acoustic \gls{slam} offers a solution to automate the involved procedures, and it has been proposed for sound source localization \cite{acousticSLAM} and room shape determination \cite{acoustic_roomshape}. In real life and the Techtile environment, spatial-embedded sources and receivers (e.g. in walls) using non-intrusive signals (e.g. ultrasound) are preferred. The Techtile infrastructure offers the ideal flexibility to conduct research into the optimization of these acoustic \gls{slam} principles. The modular tile system enables to create different room shapes with ease.

\section{Challenges}
\label{sec:challenges}
In addition to many research opportunities, practical environments also entail challenges that need to be tackled. The most important challenges with respect to acoustic research are addressed.

\subsection{Acoustic Characteristics}
The Techtile environment can be seen as a real-life, acoustically difficult environment given the high RT60 value and ambient noise factors. Mainly the relatively high RT60 value has an influence on the accuracy and reliability of indoor positioning systems as reflections result in multi-path effects, acoustically called reverberation, which in turn can cause erroneous distance estimates. With regard to acoustic sensing, ambient noise factors can be a disturbance factor that must be filtered out. However, this is a less challenging interference factor given the acceptable \gls{spl} levels. In general, the Techtile environment is an acoustically challenging space, given the wooden construction, which can represent realistic testing for real-live situations. Since this is a research room, the acoustic characteristics will also change each time other measuring devices, adapters, etc. are placed and used.

\subsection{Synchronization}
Coherent sampling should be provided through synchronized processes for many of the approaches in the research projects discussed. Due to the relatively slow speed of sound, $\rm \mu s$ synchronization is acceptable for acoustic research in contrast to the necessary $\rm ns$ accuracy for some \gls{rf} applications. 
When only the \gls{daq} is addressed, this challenge is relatively simple, provided that the sampling is inherently present when setting some parameters. If more complex setups are required between reciprocal infrastructure elements, the synchronization is more challenging. Multiple synchronization options can be applied such as via dedicated clock sharing cabling for high-accuracy synchronization, over-the-air synchronization \cite{over-the-air-synch} and synchronization over Ethernet \cite{ethernetSynch} as all tiles are connected with an Ethernet cable. This last option is interesting as this synchronization is already provided via the IEEE 1588v2 \gls{ptp} which offers sub-microsecond accuracy depending on the network. This accuracy can suffice for most acoustic realizations with the added advantage that it can be applied scalable and at low-cost. synchronization between the \gls{daq} and other infrastructure elements is still an open challenge.

\subsection{(Self-)calibration}
Calibration of processing and sensing units needs to be maintained. In addition to the supplied calibration of the existing devices, calibration must also be provided for in-house designed hardware, for example to map correct \gls{spl} levels for a certain designed microphone module. Since this can be time-consuming and is undesirable in consumer markets, self-calibration methods should be considered. Deployment in reality should be simplified in this way, for example on the basis of a simple cycle that the user must go through from which the needed calibration can be provided. \gls{slam} techniques can be combined with self-calibration and indoor positioning fingerprinting values can be mapped automatically, from which a model can be trained via e.g. a simulator.

\subsection{Scalability}
After using the environment for rapid prototyping, research, testing and development, a novel design of the positioning or sensing system should be  deployed in a wide range of application scenarios. Certain applications require a vast network of devices that can be scaled to the implementation's needs. This infrastructure helps in the design process as many concepts can be tested and reused in a simple way before switching to a more complicated, scalable, and commercialized setup. For example, the tiles themselves can, in certain places, be integrated in a building.

\section{Conclusion}
\label{sec:concl}
We have presented a versatile testbed, called Techtile, which can be used for a wide range of research domains. The emphasis in this paper was on distributed acoustic indoor positioning and sensing applied via this measurement infrastructure in which, in addition to the available equipment and possibilities, the acoustic characteristics of the research infrastructure were determined. Considering the test bed consists of a wooden construction, this is an acoustically challenging environment with a relatively high reverberation and a number of ambient noise components that can have a negative influence on the accuracy and reliability of acoustic systems. This real-life environment offers a very flexible nature where research can be carried out quickly in a tidy and controlled manner. We welcome researchers and developers to perform experiments in this testbed, and to reuse our designs ranging from the construction to the electronics deployed therein.

\section*{Acknowledgments}

The realization of the Techtile infrastructure is made possible thanks to the generous support of a bequest granted to the Science, Engineering and Technology Group of the KU Leuven. This research was partly funded by Research Foundation - Flanders (FWO) under grant agreement No. 90D3819N. We received hardware from Niko and On Semiconductor and thank our equipment suppliers National Instruments, Dell and W\"urth Electronics for their support for this pioneering development. Furthermore, the colleagues of the Dramco research group have made an enormous contribution by building, devising and developing Techtile.

{\footnotesize \printbibliography}

\end{document}

%% file: abbreviations.tex
\newacronym{daq}{DAQ}{data acquisition system}
\newacronym{poe}{PoE}{power-over-Ethernet}
\newacronym{usrp}{USRP}{universal software radio peripheral}
\newacronym{mems}{MEMS}{micro-electro-mechanical systems}
\newacronym{adc}{ADC}{analog to digital converter}
\newacronym{dac}{DAC}{digital to analog converter}
\newacronym{fs}{FS}{full scale}
\newacronym{rfpt}{RFPT}{radio frequency power transmission}
\newacronym{soc}{SoC}{system-on-chip}
\newacronym{sdr}{SDR}{software-defined radio}
\newacronym{lidar}{LIDAR}{laser imaging detection and ranging}
\newacronym{dds}{DDS}{direct digital synthesis}
\newacronym{rt}{RT}{reverberation time}
\newacronym{rta}{RTA}{real-time analyzer}
\newacronym{spl}{SPL}{sound pressure level}
\newacronym{tp}{TP}{test point}
\newacronym{tdoa}{TDoA}{time difference of arrival}
\newacronym{slam}{SLAM}{simultaneous localization and mapping}
\newacronym{ptp}{PTP}{precision time protocol}
\newacronym{nlos}{NLoS}{non-line-of-sight}
\newacronym{wpt}{WPT}{wireless power transfer}
\newacronym{rf}{RF}{radio frequency}

%% file: tables/accuracy_daq.tex

{\renewcommand{\arraystretch}{1.5} 
\begin{table}[h]
\centering
    \begin{tabularx}{\linewidth}{>{\hsize=.38\hsize}L >{\hsize=.11\hsize}Y >{\hsize=.11\hsize}Y>{\hsize=.11\hsize}Y>{\hsize=.11\hsize}Y>{\hsize=.11\hsize}Y>{\hsize=.11\hsize}Y}
        \hline
        
        \multicolumn{1}{l}{} & \multicolumn{4}{c}{\textbf{Analog Input}} & \multicolumn{2}{|c}{\textbf{Analog Output}} \\
        \hline
                              
         \gls{fs} range ($\rm V$) & $\pm 10$ & $\pm 5$ & $\pm 2$ & $\pm 1$ & \multicolumn{1}{|c}{$\pm 10$} & $\pm 5$\\
         Abs. Accuracy ($\rm \mu V$) & 2688 & 1379 & 654 & 313 & \multicolumn{1}{|c}{3256} & 1616 \\                          
        \hline
    \end{tabularx}
    \caption{Absolute accuracy of the analog I/O channels for different \gls{fs} ranges according to the datasheet\protect\footnotemark of the NI PXIe 6358.}
	\label{tab:resolutionDaQ}
\end{table}
}
\vspace*{-3mm}
\footnotetext{\url{https://www.ni.com/pdf/manuals/374453c.pdf}}

%% file: tables/rt60_table.tex
{\renewcommand{\arraystretch}{1.5} 
\begin{table}[h]
\centering
    \begin{tabularx}{\linewidth}{>{\hsize=.30\hsize}L >{\hsize=.1\hsize}Y >{\hsize=.1\hsize}Y>{\hsize=.1\hsize}Y>{\hsize=.1\hsize}Y>{\hsize=.1\hsize}Y>{\hsize=.1\hsize}Y>{\hsize=.1\hsize}Y}
        \hline
        
        \multicolumn{1}{l}{Frequency ($\rm Hz$)} & \multicolumn{1}{c}{\textbf{125}} & \multicolumn{1}{c}{\textbf{250}} & \multicolumn{1}{c}{\textbf{500}}& \multicolumn{1}{c}{\textbf{1000}}& \multicolumn{1}{c}{\textbf{2000}}& \multicolumn{1}{c}{\textbf{4000}} & \multicolumn{1}{c}{\textbf{8000}}\\
        \hline
                              
         RT60 ($\rm s$)         &  0.63 & 0.84  & 1.17 & 0.97  & 0.70  & 0.62  & 0.41\\
         Uncertainty ($\rm \%$) &  14.3 & 8.8  & 5.3  & 4.1  & 3.4  & 2.5  & 2.2 \\                          
        \hline
    \end{tabularx}
    \caption{Measured average RT60 values within Techtile.}
	\label{tab:rt60}
\end{table}
}
\vspace*{-3mm}